# EMMIXcskew: an R Package for the Fitting of a Mixture of Canonical Fundamental Skew $t$-Distributions

**Sharon X. Lee**
University of Queensland

**Geoffrey J. McLachlan**
University of Queensland

## Abstract

This paper presents an R package **EMMIXcskew** for the fitting of the canonical fundamental skew $t$-distribution (CFUST) and finite mixtures of this distribution (FM-CFUST) via maximum likelihood (ML). The CFUST distribution provides a flexible family of models to handle non-normal data, with parameters for capturing skewness and heavy-tails in the data. It formally encompasses the normal, $t$, and skew-normal distributions as special and/or limiting cases. A few other versions of the skew $t$-distributions are also nested within the CFUST distribution.

In this paper, an Expectation-Maximization (EM) algorithm is described for computing the ML estimates of the parameters of the FM-CFUST model, and different strategies for initializing the algorithm are discussed and illustrated. The methodology is implemented in the **EMMIXcskew** package, and examples are presented using two real datasets.

The **EMMIXcskew** package contains functions to fit the FM-CFUST model, including procedures for generating different initial values. Additional features include random sample generation and contour visualization in 2D and 3D.

## 1. Introduction

Finite mixture models, in particular normal mixture models, have been widely used in statistics and a diverse range of applied fields such as bioinformatics, biomedicine, economics, finance, genetics, image analysis, psychometrics, and social science. They provide a powerful and flexible tool for the probabilistic modelling of data, with applications ranging from density estimation to clustering, classification, and discriminant analysis. For a survey on mixture models and their applications, see Everitt and Hand (1981), Titterington *et al.* (1985), McLachlan and Basford (1988), Lindsay (1995), Böhning (2000), McLachlan and Peel (2000), and Frühwirth-Schnatter (2006), the edited volume of Mengersen *et al.* (2011), and also the papers by Banfield and Raftery (1993) and Fraley and Raftery (1998).

In recent years, mixture models with skew component distributions have received increasing attention. These models adopt densities that can take more flexible distributional shapes than the traditional normal and *t*-distributions as component distributions, rendering them suitable for a wider range of applications. Of these, the skew *t*-distribution is gaining popularity due to its ability to handle both the asymmetry and heavy-tailedness in the data. In particular, a number of different formulations of skew *t*-distribution have been used extensively in the model-based clustering literature (see, for example, Lee and McLachlan (2014a, 2016a), Azzalini *et al.* (2016), McLachlan and Lee (2016) and the references therein). They have also found many applications in a range of fields, including astrophysics (Riggi and Ingrassia 2013), financial risk analysis and modelling (Soltyk and Gupta 2011, Bernardi 2013, Lee and McLachlan 2013b, Abanto-Valle *et al.* 2015), fisheries science (Contreras-Reyes and Arellano-Valle 2013), flow cytometry (Pyne *et al.* 2009, Frühwirth-Schnatter and Pyne 2010, Rossin *et al.* 2011, Ho *et al.* 2012, Hu *et al.* 2013, Pyne *et al.* 2014, Lee *et al.* 2014, Lin *et al.* 2016, 2015, Lee *et al.* 2016c, Pyne *et al.* 2015), image segmentation (Lee and McLachlan 2013a), pharmaceutical science (Schaarschmidt *et al.* 2015), and the social sciences (Muthén and Asparouhov 2014, Asparouhov and Muthén 2016). For a comprehensive survey of skew distributions, see, for example, the articles by Azzalini (2005), Arellano-Valle and Azzalini (2006), Arellano-Valle *et al.* (2006), the book edited by Genton (2004), and the recent monograph by Azzalini and Capitanio (2014).

Recently, Lee and McLachlan (2016a) introduced a finite mixture of canonical fundamental skew *t* (FM-CFUST) distribution. This formulation of the skew *t*-distribution has a general $p \times q$ matrix of skewness of parameters (Arellano-Valle and Genton 2005). It thus provides a more general characterisation than the restricted and unrestricted skew *t*-distributions (adopting the terminology of Lee and McLachlan (2013c)). This paper describes an R package **EMMIXcskew** for the fitting of the FM-CFUST model. It implements the EM algorithm described in Lee and McLachlan (2016a) and provides other functionalities such as random sample generation, density evaluation, and the plotting of contours in 2D and 3D.

The remainder of this paper is organized as follows. Section 2 provides a brief description of the CFUST distribution and its nested models. Section 3 outlines an EM algorithm for fitting finite mixtures of CFUST distributions and examines different approaches for generating starting values for this EM algorithm. In the next two sections, the usage of the **EMMIXcskew** package is illustrated using real and simulated examples. Finally, we conclude with some brief remarks in Section 6.

## 2. The CFUST and related distributions

To establish notation, let $\boldsymbol{Y}$ be $p$-dimensional random vector that follows a multivariate CFUST distribution, denoted by $\boldsymbol{Y} \sim CFUST_{p,q}(\boldsymbol{\mu}, \boldsymbol{\Sigma}, \boldsymbol{\Delta}, \nu)$. Then the density of $\boldsymbol{Y}$ is given by

$$f(\boldsymbol{y}; \boldsymbol{\mu}, \boldsymbol{\Sigma}, \boldsymbol{\Delta}, \nu) \;=\; 2^q \, t_p(\boldsymbol{y}; \boldsymbol{\mu}, \boldsymbol{\Omega}, \nu) \; T_q\left(\boldsymbol{c}(\boldsymbol{y})\sqrt{\frac{\nu+p}{\nu+d(\boldsymbol{y})}}; \boldsymbol{0}, \boldsymbol{\Lambda}, \nu+p\right), \qquad (1)$$

where

$$\boldsymbol{\Omega} = \boldsymbol{\Sigma} + \boldsymbol{\Delta}\boldsymbol{\Delta}^{\top},$$
$$\boldsymbol{c}(\boldsymbol{y}) = \boldsymbol{\Delta}^{\top}\boldsymbol{\Omega}^{-1}(\boldsymbol{y} - \boldsymbol{\mu}),$$
$$\boldsymbol{\Lambda} = \boldsymbol{I}_q - \boldsymbol{\Delta}^{\top}\boldsymbol{\Omega}^{-1}\boldsymbol{\Delta},$$
$$d(\boldsymbol{y}) = (\boldsymbol{y} - \boldsymbol{\mu})^{\top}\boldsymbol{\Omega}^{-1}(\boldsymbol{y} - \boldsymbol{\mu}).$$

It can observed from (1) above that the CFUST distribution is described by the parameters $(\boldsymbol{\mu}, \boldsymbol{\Sigma}, \boldsymbol{\Delta}, \nu)$, where $\boldsymbol{\mu}$ is a $p$-dimensional vector of location parameters, $\boldsymbol{\Sigma}$ is a positive definite scale matrix, $\boldsymbol{\Delta}$ is a $p \times q$ matrix of skewness parameters, and $\nu$ is a scalar degrees of freedom that regulate the tails of the distribution. In the above, we let $t_p(\boldsymbol{y}; \boldsymbol{\mu}, \boldsymbol{\Omega}, \nu)$ denote the $p$-dimensional $t$-distribution with location parameter $\boldsymbol{\mu}$, scale matrix $\boldsymbol{\Omega}$, and degrees of freedom $\nu$, and $T_q(.)$ is the $q$-dimensional (cumulative) $t$-distribution function.

As mentioned previously, the CFUST distribution includes some commonly used distributions as special and/or limiting cases. Taking $\boldsymbol{\Delta} = \boldsymbol{0}$ reduces (1) to the symmetric multivariate $t$-density $t_p(\boldsymbol{\mu}, \boldsymbol{\Omega}, \nu)$, and further letting $\nu \to \infty$ and $\nu = 1$ leads to the multivariate normal $N_p(\boldsymbol{\mu}, \boldsymbol{\Omega})$ and Cauchy $C_p(\boldsymbol{\mu}, \boldsymbol{\Omega})$ distributions, respectively. If $\boldsymbol{\Delta}$ is constrained to be a diagonal matrix, we obtain the skew $t$-distribution of Sahu *et al.* (2003) which is referred to as the unrestricted skew $t$-distribution using the terminology in Lee and McLachlan (2014a, 2013c). To obtain the classical skew $t$-distribution by Azzalini and Capitanio (2003) from (1), one can set $q = 1$ or take $\boldsymbol{\Delta}$ to be a matrix of zeros except for one column (Lee and McLachlan 2016a). This formulation of the skew $t$-distribution, referred to as the restricted skew $t$-distribution, is equivalent to that given by Branco and Dey (2001), Gupta (2003), Lachos *et al.* (2010) and Pyne *et al.* (2009); see Lee and McLachlan (2013c). Analogously, the restricted and unrestricted skew normal distributions can be obtained by placing appropriate constraints on $\boldsymbol{\Delta}$ and letting $\nu \to \infty$. Some properties of the CFUST distribution are described in Arellano-Valle and Genton (2005). It is of interest to note that this distribution suffers an identifiability issue as discussed in Lee and McLachlan (2016). In brief, this means that the CFUST density is invariant under permutations of the columns of the skewness matrix $\boldsymbol{\Delta}$, but this does not affect parameter estimation. Hence, in practice, the user only needs to be aware that changing the order of the columns of $\boldsymbol{\Delta}$ does not affect the density of the CFUST model.

There are several R packages on CRAN that deal with (multivariate) mixture models with skewed component densities. In particular, the restricted and unrestricted versions of skew $t$-mixture models are implemented in **EMMIXskew** (Wang *et al.* 2009) and **EMMIXuskew** (Lee and McLachlan 2013d), respectively. The package **mixsmsn** (Prates *et al.* 2013) implements the family of finite mixtures of scale-mixture of skew-normal distributions, which includes a skew normal distribution and a skew $t$-distribution that is equivalent to the restricted skew normal distribution and restricted skew $t$-distribution, respectively. However, the estimation procedure used in **mixsmsn** imposes the condition that all components of the skew $t$-mixture model share a common value for the degrees of freedom. A recently developed package **MixGHD** (Tortora *et al.* 2015) provides functions to fit finite mixtures of generalized hyperbolic distributions. For the classical multivariate skew-normal and skew $t$-distributions, the **sn** package (Azzalini 2014) can be used. For traditional normal mixture model and related tools, a number of other packages are available on CRAN, such as **bgmm** (Biecek *et al.* 2012), **flexmix** (Leisch 2004, Grün and Leisch 2008), **mclust** (Fraley and Raftery 2007, Scrucca *et al.* 2016), and **mixtools** (Benaglia *et al.* 2009).

# 3. Fitting mixtures of CFUST distributions via the EM algorithm

The density of a finite mixture model is given by a convex linear combination of component densities. More formally, adopting the CFUST distribution as component densities, we obtain a finite mixture of CFUST (FM-CFUST) distribution with density given by

$$f(\boldsymbol{y}; \boldsymbol{\Psi}) = \sum_{h=1}^{g} \pi_h f(\boldsymbol{y}; \boldsymbol{\mu}_h, \boldsymbol{\Sigma}_h, \boldsymbol{\Delta}_h, \nu_h), \tag{2}$$

where $\pi_h$ $(h = 1, \ldots, g)$ are the mixing proportions and $f(.)$ denotes the CFUST density given by (1). The mixing proportions satisfy $\pi_h \geq 0$ and $\sum_{h=1}^{g} \pi_h = 1$. The vector $\boldsymbol{\Psi} = (\pi_1, \ldots, \pi_{g-1}, \boldsymbol{\theta}_1^T, \ldots, \boldsymbol{\theta}_g^T)$ contains all the unknown parameters of the model, with $\boldsymbol{\theta}_h$ containing the elements of $\boldsymbol{\mu}_h$ and $\boldsymbol{\delta}_h$, the distinct elements of $\boldsymbol{\Sigma}_h$, and $\nu_h$.

For the fitting of the FM-CFUST model, we employ the EM algorithm (Dempster *et al.* 1977) to compute the maximum likelihood (ML) estimate of the parameters of the model. The EM algorithm proceeds by alternating repeatedly between the E- and M-steps until the changes in the log likelihood values are less than some specified small value in the case of convergence.

To facilitate parameter estimation via the EM algorithm, a set of latent variables are introduced, namely the component labels $\boldsymbol{Z}_j$ (corresponding to the $j = 1, \ldots, n$ independent observations on $\boldsymbol{Y}$), alongside two random variables $\boldsymbol{U}_j$ and $W_j$ that follow a half-normal distribution and a gamma distribution, respectively. Thus, the complete-data for the FM-CFUST model consist of these missing variables and the observations $\boldsymbol{y}_j$. This leads to a four-level hierarchical characterization of the FM-CFUST model, given by

$$\begin{aligned}
\boldsymbol{Y}_j \mid \boldsymbol{u}_j, w_j, z_{hj} = 1 &\sim N_p\left(\boldsymbol{\mu} + \boldsymbol{\Delta}_h \boldsymbol{u}_j, \frac{1}{w_j}\boldsymbol{\Sigma}_h\right), \\
\boldsymbol{U}_j \mid w_j, z_{hj} = 1 &\sim HN_q\left(\boldsymbol{0}, \frac{1}{w_j}\boldsymbol{I}_q\right), \\
W_j \mid z_{hj} = 1 &\sim \text{gamma}\left(\frac{\nu_h}{2}, \frac{\nu_h}{2}\right), \\
\boldsymbol{Z}_j &\sim \text{Mult}_g\left(1, \boldsymbol{\pi}\right),
\end{aligned} \tag{3}$$

where $\boldsymbol{Z}_j$ is a $g$-dimensional vector of binary component labels such that $Z_{hj} = (\boldsymbol{Z}_j)_h = 1$ if the $j$th observation belongs to the $h$th component and zero otherwise. Here, $HN_q(\boldsymbol{0}, \boldsymbol{\Sigma})$ denotes the $q$-dimensional half-normal distribution with scale matrix $\boldsymbol{\Sigma}$, gamma$(\alpha, \beta)$ denotes the gamma distribution with mean $\frac{\alpha}{\beta}$, and $\text{Mult}_g(1, \boldsymbol{\pi})$ denotes the multinomial distribution with one draw and $g$ categories with probabilities specified by $\boldsymbol{\pi}$. We now outline the E- and M-steps of the EM algorithm for fitting the FM-CFUST model.

## 3.1. The E-step

The E-step of the EM algorithm requires the calculation of the so-called $Q$-function, $Q(\boldsymbol{\Psi}; \boldsymbol{\Psi}^{(k)})$, which is the conditional expectation of the complete-data log likelihood given the observed data $\boldsymbol{y}$, using the current estimate of $\boldsymbol{\Psi}$, which is denoted by $\boldsymbol{\theta}^{(k)}$ after the $k$th iteration. It follows that on the $(k+1)$th iteration, the E-step requires the following five conditional expectations to be calculated,

$$
\begin{align}
z_{hj}^{(k)} &= E_{\Psi^{(k)}}\left[z_{hj} = 1 \mid \boldsymbol{y}_j\right], \tag{4}\\
w_{hj}^{(k)} &= E_{\Psi^{(k)}}\left[w_{hj} \mid \boldsymbol{y}_j, z_{hj} = 1\right], \tag{5}\\
e_{1hj}^{(k)} &= E_{\Psi^{(k)}}\left[\log(w_{hj}) \mid \boldsymbol{y}_j, z_{hj} = 1\right], \tag{6}\\
\boldsymbol{e}_{2hj}^{(k)} &= E_{\Psi^{(k)}}\left[w_{hj}\boldsymbol{u}_{hj} \mid \boldsymbol{y}_j, z_{hj} = 1\right], \tag{7}\\
\boldsymbol{e}_{3hj}^{(k)} &= E_{\Psi^{(k)}}\left[w_{hj}\boldsymbol{u}_{hj}\boldsymbol{u}_{hj}^{\top} \mid \boldsymbol{y}_j, z_{hj} = 1\right]. \tag{8}
\end{align}
$$

The expressions for (4) to (8) are given in Lee and McLachlan (2016a) and therefore not repeated here. However, it should be noted that $e_{1hj}^{(k)}$ can be evaluated using different approaches, two of which are described in the above reference. For simplicity, the **EMMIXcskew** package implements the one-step-late (OSL) approach for this conditional expectation. It should be noted that the use of the approximate OSL approach to calculate $e_{1hj}^{(k)}$ can result in the incomplete-data likelihood not increasing monotonically. This conditional expectation can be calculated more accurately by a power series derived in Lee and McLachlan (2014b, 2016a) for which monotonicity of the likelihood is preserved. An implementation of this option will be provided in a future update of this package.

### 3.2. The M-step

On the $(k + 1)$th iteration of the the M-step, the current estimate of $\boldsymbol{\Psi}$, $\boldsymbol{\Psi}^{(k)}$, is updated to $\boldsymbol{\Psi}^{(k+1)}$, which is chosen to globally maximize $Q(\boldsymbol{\Psi}; \boldsymbol{\Psi}^{(k)})$ over $\boldsymbol{\Psi}$. For the FM-CFUST model, the M-step leads to the following updates:

$$
\begin{align}
\pi_h^{(k+1)} &= \frac{1}{n}\sum_{j=1}^{n} z_{hj}^{(k)}, \\
\boldsymbol{\mu}_h^{(k+1)} &= \frac{\sum_{j=1}^{n} z_{hj} w_{hj}^{(k)} \boldsymbol{y}_j - \boldsymbol{\Delta}_h^{(k)} \sum_{j=1}^{n} z_{hj}^{(k)} \boldsymbol{e}_{2hj}^{(k)}}{\sum_{j=1^n} z_{hj}^{(k)} w_{hj}^{(k)}}, \\
\boldsymbol{\Delta}_h^{(k+1)} &= \left[\sum_{j=1}^{n} z_{hj}^{(k)}\left(\boldsymbol{y}_j - \boldsymbol{\mu}_h^{(k+1)}\right)\boldsymbol{e}_{2hj}^{(k)\top}\right]\left[\sum_{j=1}^{n} z_{hj}^{(k)}\boldsymbol{e}_{3hj}^{(k)}\right]^{-1}, \tag{9}\\
\boldsymbol{\Sigma}_h^{(k+1)} &= \left\{\sum_{j=1}^{n} z_{hj}^{(k)}\left[w_{hj}^{(k)}\left(\boldsymbol{y}_j - \boldsymbol{\mu}_h^{(k+1)}\right)\left(\boldsymbol{y}_j - \boldsymbol{\mu}_h^{(k+1)}\right)^T\right.\right. \\
&\quad \left. -\boldsymbol{\Delta}_h^{(k+1)}\boldsymbol{e}_{2hj}^{(k)}\left(\boldsymbol{y}_j - \boldsymbol{\mu}_h^{(k+1)}\right)^{\top} - \left(\boldsymbol{y}_j - \boldsymbol{\mu}_h^{(k+1)}\right)\boldsymbol{e}_{2hj}^{(k)\top}\boldsymbol{\Delta}_h^{(k+1)\top}\right. \\
&\quad \left.\left. +\boldsymbol{\Delta}_h^{(k+1)}\boldsymbol{e}_{3hj}^{(k)\top}\boldsymbol{\Delta}_h^{(k+1)\top}\right]\right\}\left[\sum_{j=1}^{n} z_{hj}^{(k)}\right]^{-1}. \tag{10}
\end{align}
$$

An update of the degrees of freedom $\nu_h$ is obtained by solving the following equation for

$\nu_h^{(k+1)}$,

$$0 = \left(\sum_{h=1}^n z_{hj}^{(k)}\right)\left[\log\left(\frac{\nu_h^{(k+1)}}{2}\right) - \psi\left(\frac{\nu_h^{(k+1)}}{2}\right) + 1\right] - \sum_{j=1}^n z_{hj}^{(k)}\left(e_{1hj}^{(k)} - w_{hj}^{(k)}\right),$$

where $\psi(\cdot)$ denotes the digamma function.

### 3.3. Generating initial values for parameters

As the log likelihood function may exhibit a complicated profile with many local maxima and the EM algorithm is sensitive to its initial values, it is important to choose good starting values. In this section, we consider three strategies for generating valid initial values for the EM algorithm for the FM-CFUST model. For the remainder of this section, we suppress the subscript $h$ (denoting the index of a component in a mixture model) for notational convenience.

*Nested approach*

An intuitive approach is to start the EM algorithm with the solution given by one of the nested models of a CFUST distribution, for example, the results from fitting a normal or $t$-mixture model. This option is available in **EMMIXcskew** with the `fmcfust.init` function (see Section 5.2), which accepts the outputs from the packages **EMMIXskew** and **EMMIXuskew**. The former package provide routines to fit mixtures of (multivariate) normal, $t$-, (restricted) skew normal, and skew $t$-distributions, whereas the latter package fits a mixture of unrestricted skew $t$-distributions.

*Method of moments-based approach*

Another approach is based on the moments of an unrestricted multivariate skew normal (uMSN) distribution. As noted earlier, the uMSN distribution is a nested case of the CFUST distribution. It can be characterized as the convolution of a truncated normal variable and a multivariate normal variable as follows,

$$\boldsymbol{Y}_j = \boldsymbol{\mu} + \boldsymbol{\Delta}|\boldsymbol{U}_{0j}| + \boldsymbol{U}_{1j}, \tag{11}$$

where $\boldsymbol{\mu} \in \mathbb{R}$, $\boldsymbol{\Delta} = \text{diag}(\boldsymbol{\delta})$ is a diagonal matrix of skewness parameters with diagonal elements given by $\boldsymbol{\delta}$, $\boldsymbol{U}_{0j} \sim N_p(\boldsymbol{0}, \boldsymbol{I}_p)$ and $\boldsymbol{U}_{1j} \sim N_p(\boldsymbol{0}, \boldsymbol{\Sigma})$. The uMSN distribution (11) has mean and variance given by

$$E(\boldsymbol{Y}_j) = \boldsymbol{\mu} + \sqrt{\frac{2}{\pi}}\boldsymbol{\delta},$$

and

$$\text{cov}(\boldsymbol{Y}_j) = \boldsymbol{\Sigma} + (1 - \frac{2}{\pi})\boldsymbol{\Delta}^2, \tag{12}$$

respectively. On rearranging the above expressions, we obtain an expression for $\boldsymbol{\mu}$ and $\boldsymbol{\Sigma}$ in terms of $\boldsymbol{\delta}$ (recall that by definition $\boldsymbol{\Delta} = \text{diag}(\boldsymbol{\delta})$ for the uMSN distribution). To obtain an initial value for $\boldsymbol{\delta}^{(0)}$, one can consider reducing the values of the diagonal elements of $\boldsymbol{\Sigma}^{(0)}$ by an arbitrary proportion $(1 - a)$ where $a \in [0, 1]$ (Lin 2010). This leads a set of expressions given by

$$
\begin{aligned}
\boldsymbol{\delta}^{(0)} &= \pm\sqrt{\frac{\pi(1-a)}{\pi-2}}\boldsymbol{s}^*, \\
\boldsymbol{\Sigma}^{(0)} &= \boldsymbol{S} + (a-1)\operatorname{diag}(\boldsymbol{s}^*), \\
\boldsymbol{\mu}^{(0)} &= \bar{\boldsymbol{y}} - \sqrt{\frac{2}{\pi}}\boldsymbol{\delta}^{(0)},
\end{aligned}
\tag{13}
$$

where the sign of each element of $\boldsymbol{\delta}^{(0)}$ is given by the sign of the third-order sample moment of the corresponding variable about its sample mean. In (13) above, $\boldsymbol{s}^*$ is a $p$-dimensional vector containing the diagonal elements of the sample covariance matrix $\boldsymbol{S}$, and $\bar{\boldsymbol{y}}$ denotes the sample mean. Concerning the degrees of freedom, it can be set (initially) to a large number to reflect a uMSN distribution.

*Transformation approach*

A third approach is based on a transformation of $\boldsymbol{Y}_j$ in an attempt to better handle the correlation of the variables in $\boldsymbol{Y}_j$. We consider the transformation vector $\boldsymbol{X}_j = \boldsymbol{C}\boldsymbol{Y}_j$, where $\boldsymbol{C}$ is an orthogonal matrix such that the covariance matrix of $\boldsymbol{X}_j$, $\operatorname{cov}(\boldsymbol{X}_j)$, is diagonal. In practice, we work with the sample covariance matrix of $\boldsymbol{Y}_j$. Then we can fit a uMST distribution to the transformed vector $\boldsymbol{X}_j$, where each $\boldsymbol{X}_j$ can be characterized as

$$
\begin{aligned}
\boldsymbol{X}_j &= \boldsymbol{C}\boldsymbol{Y}_j \\
&= \boldsymbol{\mu} + \boldsymbol{\Delta}|\boldsymbol{U}_{0j}| + \boldsymbol{U}_{1j},
\end{aligned}
\tag{14}
$$

where $\boldsymbol{U}_{0j}$ and $\boldsymbol{U}_{1j}$ follow a central multivariate $t$-distribution with $\nu$ degrees of freedom and scale matrix given by $\boldsymbol{I}_q$ and $\boldsymbol{\Sigma}$, respectively. Note that in (14) above, we have used a stochastic representation of the CFUST distribution that is analogous to (11) for a uMSN distribution. On pre-multiplying $\boldsymbol{X}_j$ by $\boldsymbol{C}^\top$ in (14) to obtain $\boldsymbol{Y}_j$, we have

$$
\boldsymbol{Y}_j = \boldsymbol{C}^\top\boldsymbol{\mu} + \boldsymbol{C}^\top\boldsymbol{\Delta}|\boldsymbol{U}_{0j}| + \boldsymbol{C}^\top\boldsymbol{U}_{1j}.
\tag{15}
$$

This suggests that an initial value for $\boldsymbol{\mu}$ and for $\boldsymbol{\Delta}$ in a CFUST distribution can be given by $C^\top\hat{\boldsymbol{\mu}}$ and $C^\top\hat{\boldsymbol{\Delta}}$, respectively, where $\hat{\boldsymbol{\mu}}$ and $\hat{\boldsymbol{\Delta}}$ are the estimates of $\boldsymbol{\mu}$ and $\boldsymbol{\Delta}$ obtained by fitting the uMST distribution to the $\boldsymbol{X}_j$. However, it should be noted that if the true distribution of $\boldsymbol{Y}_j$ were a CFUSN distribution, then the transformed data $\boldsymbol{X}_j$ may not necessarily have a uMSN distribution even though $\operatorname{cov}(\boldsymbol{X}_j)$ is diagonal. This happens when the off-diagonal elements of $\boldsymbol{\Sigma}$ cancel out the off-diagonal elements of $\boldsymbol{\Delta}\boldsymbol{\Delta}^T$. But it might be expected that in most situations where the sample covariance matrix of the $\boldsymbol{X}_j$ is approximately diagonal, the matrix $\boldsymbol{\Sigma}$ and the skewness matrix $\boldsymbol{\Delta}$ are both diagonal or close to it.

In the case where a mixture of CFUST distributions is to be fitted rather than a single component distribution, we would need to first cluster the $\boldsymbol{Y}_j$ into $g$ clusters and proceed separately within each cluster as described above.

The above three methods are implemented in the function `fmcfust.init` of the **EMMIXcskew** package. By default, it adopts the moments method. An example on a real dataset demonstrating the use of these approaches is given in Section 5.2.

### 3.4. Stopping criterion

We adopt the Aitken acceleration-based stopping criterion as the default strategy to assess the convergence of the EM algorithm for the FM-CFUST model. The **EMMIXcskew** package also provides a few other criteria as an alternative, including those based on the relative change in the log likelihood value and estimates of the parameters of the model.

*Aiken acceleration-based approach*

In brief, when using the default strategy, our algorithm terminates when the absolute difference between a log likelihood value and its asymptotic estimate is smaller than a specified tolerance, $\epsilon$; that is, when

$$\left| \ell_\infty^{(k+1)} - \ell^{(k+1)} \right| < \epsilon, \tag{16}$$

where $\ell^{(k+1)}$ is the log likelihood value at the $(k+1)$th iteration and $\ell_\infty^{(k+1)}$ is its asymptotic estimate, given by

$$\ell_\infty^{(k+1)} = \ell^{(k)} + \frac{\ell^{(k+1)} - \ell^{(k)}}{1 - a^{(k)}}. \tag{17}$$

In the above, $a^{(k)} = \frac{\ell^{(k+1)} - \ell^{(k)}}{\ell^{(k)} - \ell^{(k-1)}}$ denotes the Aitken's acceleration at the $k$th iteration (Böhning *et al.* 1994, McLachlan and Krishnan 1997). The default tolerance of $\epsilon = 10^{-6}$ is applied to the examples in the following sections, but the user can specify a different value.

*Relative likelihood-based approach*

Another commonly used stopping criterion is to monitor the relative changes in the log likelihood values at the end of each iteration and to stop the algorithm when the (relative) difference between two successive log likelihood values is less than a specified threshold. More formally, our algorithm terminates when

$$\frac{\left| \ell^{(k+1)} - \ell^{(k)} \right|}{\left| \ell^{(k+1)} \right|} < \epsilon, \tag{18}$$

where the threshold $\epsilon$ is set to $10^{-6}$ by default. Again, the user can specify a different threshold.

*Relative parameters-based approach*

Apart from tracking the changes in the log likelihood value, one can also monitor the changes in the parameter estimates. In this case, the algorithm is considered to have converged when the relative change in all the parameter estimates is less than a specified threshold $\epsilon$. Note that this criterion implies that the relative changes of all the free parameters needs to be smaller than $\epsilon$, that is, all elements of $\boldsymbol{\Psi}$ including the mixing proportions. Thus, the EM algorithm terminates when

$$\frac{\left| \boldsymbol{\Psi}^{(k+1)} - \boldsymbol{\Psi}^{(k)} \right|}{\left| \boldsymbol{\Psi}^{(k+1)} \right|} < \epsilon. \tag{19}$$

Again, the default tolerance is $\epsilon = 10^{-6}$.

### 3.5. Notes on the EM implementation

In the EM algorithm described in this paper, the number of components $g$ and the dimension $q$ of the skewing vector must be specified. In practice, these are typically unknown and model selection criteria are employed to aid in choosing an appropriate value of $g$ and $q$. Some of the more commonly used information criterion include the Bayesian information criterion (BIC; Schwarz (1978)), given by

$$BIC = m \log n - 2\ell, \tag{20}$$

and the Akaike information criterion (AIC; Akaike (1974)), given by

$$BIC = 2m - 2\ell, \tag{21}$$

where $m$ is the number of free parameters, $n$ is the number of observations, and $\ell$ is the value of the log likelihood function at the fitted parameter vector. As is typical in fitting a mixture of factor analyzers (MFA) model, one may fit the FM-CFUST model for a range of values of $p$ and $q$ and choose the combination of $p$ and $q$ corresponding to the lowest AIC or BIC. An alternative strategy for automatically selecting an appropriate $g$ was considered in Lee and McLachlan (2016c) which is based on the minimum message length (MML) criteria. However, concerning the value of $q$, it was observed in Lee and McLachlan (2016a) that when fitting the CFUST model with $q = p$ it was able to model data generated from the rMST ($q = 1$) and uMST ($q = p$ and $\Delta$ is a diagonal matrix) distributions quite well. In particular, the estimated $\Delta$ matches the true $\Delta$ reasonably well. For example, in the case of data generated from a rMST distribution, all but one of the columns of the estimated $\Delta$ has elements being relatively small, thus resembling the $q = 1$ case. Hence, in the implementation of the EM algorithm in the **EMMIXcskew** package, the default value of $q$ is set to $p$. But the user is encouraged to experiment with different values of $q$ when fitting the FM-CFUST model.

## 4. Using the EMMIXcskew package

The **EMMIXcskew** package implements the EM algorithm described in Section 3 and provides additional functions such as random sample generation, density evaluation, and graphics outputs. The software is primarily written in R. **EMMIXcskew** will be made available on the Comprehensive R Archive Network (CRAN).

In this sequel, we now demonstrate the basic usage of the **EMMIXcskew** package using simple examples. In particular, the following main routines are discussed:

- `dfmcfust`: evaluation of density values;

- `rfmcfust`: generation of a random sample from a FM-CFUST distribution;

- `fmcfust`: fitting a FM-CFUST model;

- `init.cfust`: generation of initial values for use in `fmcfust`;

- `fmcfust.contour.2d`: plotting a 2D graph of the contours of a FM-CFUST model;

| Parameters | R arguments | Dimensions | Description |
|:---:|:---:|:---:|:---:|
| $\boldsymbol{\mu}$ | `mu` | $p \times 1 \times g$ | the location parameters |
| $\boldsymbol{\Sigma}$ | `sigma` | $p \times p \times g$ | the scale matrices |
| $\boldsymbol{\Delta}$ | `delta` | $p \times q \times g$ | the skewness parameters |
| $\nu$ | `dof` | $g \times 1$ | the degrees of freedom |
| $\pi$ | `pro` | $g \times 1$ | the mixing proportions |

Table 1: Structure of the model parameters in **EMMIXcskew**.

- `fmcfust.contour.3d`: plotting 3D surface contours of a FM-CFUST model.

## 4.1. The density function of FM-CFUST distribution

The density of a FM-CFUST distribution is calculated by the `dcfust` function. The inputs to be passed into this function are similar to that in `dfmmst` from the **EMMIXuskew** package, and must be structured as described in Table 1. Briefly, the parameters $\boldsymbol{\mu}$, $\boldsymbol{\Sigma}$, and $\boldsymbol{\delta}$ are each implemented as a list of `g` matrices, where `g` is the number of components in the fitted model. Each element of the `list` objects `mu`, `sigma`, and `delta` ($h = 1, \ldots, g$) must be specified as a $p \times 1$, $p \times p$, and $p \times q$ matrix, respectively. The parameters `dof` and `pro` are $g$ by 1 arrays, representing the degrees of freedom and the mixing proportions for each component, respectively. Finally, the input data are specified by `dat`, an $n \times p$ matrix where each row represents an individual observation.

Typically, one may be interested in calculating density values for a fitted FM-CFUST model (obtained from the `fmcfust` function). In this case, the output object of the fitted model can be directly passed in to `dfmcfust` as a single argument `known`. Note that if both `known` and all the model parameters were provided by he user, only the values specified by `known` would be used. Issuing the following command will return a vector of $n \times 1$ density values.

```
dfmcfust(dat, mu, sigma, delta, dof, pro, known=NULL)
```

For a single CFUST distribution (that is, `g=1`), the `dcfust` function can be used. Here, the arguments `mu`, `sigma`, and `delta` need not be a `list` object, but `mu` can be a numeric array, and `sigma` and `delta` are matrices. Similar to the above function call, the `dcfust` can be called at the R command prompt as follows, which will return a numeric vector of density values.

```
dcfust(dat, mu, sigma, delta, dof)
```

## 4.2. Fitting a single CFUST distribution

To fit the FM-CFUST model with a single component ($g = 1$), the main routine `fmcfust` can be used. This implements the EM algorithm described in Section 3, with the default strategy for initial values given in (13). By default, $q$ is assumed to be the same as $p$. A typical call of `fmcfust` is:

```
fmcfust(g, dat, q=p, initial=NULL, known=NULL, itmax=100, eps=1e-6,
nkmeans=20, verbose=TRUE)
```

As a simple example, we consider the `iris` dataset, available directly in R. For illustration purposes, we look at the Versicolor species and focus on the two variables `Sepal.Widthh` and `Pedal.Length`. We first create a new data object `iris.versicolor` with the required data, then execute the `fmcfust` function with $g = 1$. This is the minimum information that must be supplied to `fmcfust`.

```
R> library(EMMIXcskew)
R> data(iris)
R> iris.versicolor <- iris[iris$Species=="versicolor",2:3]
R> Fit.versicolor <- fmcfust(1, iris.versicolor)
```

The above command will return a `fmcfust` object, containing the final estimate of the parameters, the predicted cluster labels, and a number of measures associated with the fitted model. The final estimated parameters can be accessed using `Fit.versicolor$mu`, `Fit.versicolor$sigma`, `Fit.versicolor$delta`, `Fit.versicolor$dof`, and `Fit.versicolor$pro`. To view these parameters, `summary` can be called:

```
R > summary(Fit.versicolor)
Finite Mixture of Multivariate CFUST Distribution
with 1 component

Mean:
           [,1]
[1,] 3.415878
[2,] 4.886890

Scale matrix:
            [,1]            [,2]
[1,]   0.006138577 -0.007283746
[2,] -0.007283746  0.020649780

Skewness matrix:
           [,1]           [,2]
[1,] -0.4901844 -0.37242352
[2,] -0.9067630  0.03643873

Degrees of freedom:
[1] 87.47343
```

The other arguments of `fmcfust` are similar to that used in `fmmst` from the **EMMIXuskew** package. Briefly, `known` is a list of model parameters that are known *a priori* and, if supplied, will not be updated in the iterations of EM algorithm. The arguments `itmax` and `eps` determine when the EM algorithm is terminated. If either the maximum number of iterations as specified by `itmax` is reached or the tolerance as specified by `eps` is obtained, the EM loop will terminate. User specified initial values can be supplied using `initial`. Note that this must be a `list` object structured as in Table 1. The argument `nkmeans` specifies the number of $k$-means trials to be preformed when using the default starting strategy.

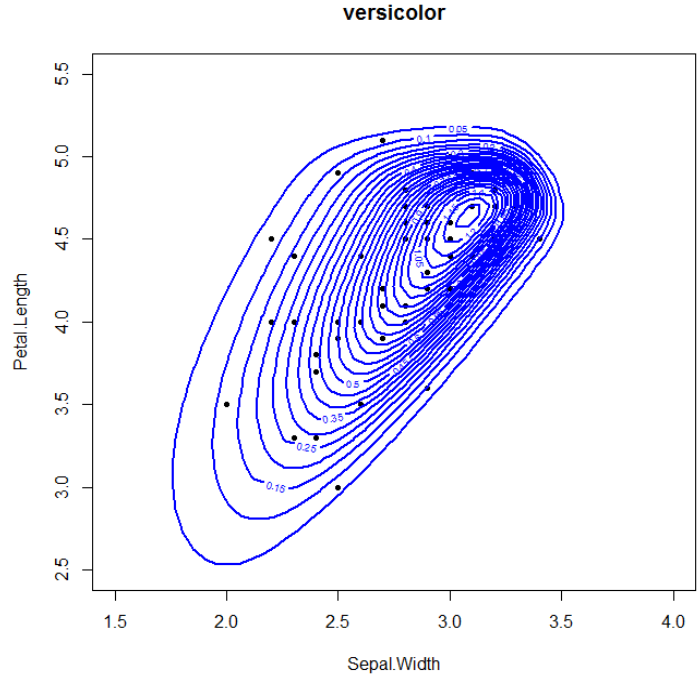

Figure 1: Contours of a FM-CFUST model fitted to the versicolor data.

With `verbose=TRUE`, `fmcfust` prints the log likelihood value at each iteration and displays a summary of the estimated parameters of the model.

Note that in the above example we have used the default starting strategy to generate initial values, and assume $q = p$. As pointed out in Section 3, this may not always give the optimal fit. It is highly recommended that the EM algorithm is run from a range of different starting values. Some alternative methods for generating different starting values are discussed in Section 3.3. These are implemented in `init.cfust` (see Section 5.2 for further discussions). In addition, the user can also experiment with different values of $q$. However, it is interesting to note that for the simulated dataset of Section 4.1 in Lee and McLachlan (2016a), it was observed that the FM-CFUST model is able to (roughly) recover the structure of $\boldsymbol{\Delta}$ without prior knowledge of any constraints on the matrix of skewness parameters.

To assist in choosing a suitable model for the data from a range of different fitted results (for example, using different starting values), log likelihood values and information measures such as AIC and BIC can be compared. These values are available as part of the output of `fmcfust` and can be accessed using `Fit.versicolor$loglik`, `Fit.versicolor$aic`, and `Fit.versicolor$bic`. They can also be viewed using the `print` command as shown below for the example above. The contours of the fitted model can be visualized using `fmcfust.contour.2d` (see Figure 1). Further details and examples of contour plots will be given in Section 5.5.

```
> print(Fit.versicolor)
Finite Mixture of Multivariate CFUST Distribution
with  1  component
```

```
... the first five elements omitted ...

$tau
      [,1] [,2] [,3] [,4] [,5] [,6] [,7] [,8] [,9] [,10] [,11]
[1,]    1    1    1    1    1    1    1    1    1     1     1

$clusters
 [1] 1 1 1 1 1 1 1 1 1 1 1 1 1 1 1 1 1 1 1 1
... the rest omitted..

$loglik
[1] -32.30904

$aic
[1] 84.61809

$bic
[1] 103.7383
```

### 4.3. Fitting a FM-CFUST distribution

Consider now the fitting of a three-component FM-CFUST model to the entire `iris` dataset. It consists of four geometric measurements on 150 observations of Iris, with 50 observations from each of three species of Iris (Setosa, Virginica, and Versicolor). The following code fits a FM-CFUST model using the results of a FM-uMST model as starting values.

```
R> fit.unrestricted <- fmmst(3,iris[,-5])
R> fit.iris <- fmcfust(3, iris[,-5], initial=fit.unrestricted, method="EMMIXuskew")
```

Again, this returns a `fmcfust` object.

```
Finite Mixture of Multivariate CFUST Distributions
with 3 components

   --------------------------------------------------

   Iteration  0 : loglik =  -171.9228
   Iteration  1 : loglik =  -170.8919
   Iteration  2 : loglik =  -170.419
   Iteration  3 : loglik =  -170.027
   Iteration  4 : loglik =  -169.6717
   Iteration  5 : loglik =  -169.3511
       ... rest omitted ...
   --------------------------------------------------

   Iteration 100: loglik = -154.561
```

```
   Component means:
          [,1]      [,2]      [,3]
   [1,] 4.8679805 6.345333 5.888617
   [2,] 3.2808574 3.066393 2.808073
   [3,] 1.4284854 4.474453 5.037550
   [4,] 0.1343135 1.317850 2.072705

   Component scale matrices:
   [[1]]
               [,1]         [,2]         [,3]          [,4]
   [1,]  0.0443155106  0.062560933 -0.001555864 -0.0009055587
   [2,]  0.0625609328  0.113034979 -0.005108048 -0.0020813146
   [3,] -0.0015558645 -0.005108048  0.004406555 -0.0001740700
   [4,] -0.0009055587 -0.002081315 -0.000174070  0.0014520658

   [[2]]
             [,1]        [,2]        [,3]        [,4]
   [1,] 0.09076830 0.025306942 0.025998408 0.019130182
   [2,] 0.02530694 0.017246048 0.006082013 0.013000612
   [3,] 0.02599841 0.006082013 0.018261259 0.007659824
   [4,] 0.01913018 0.013000612 0.007659824 0.013618464

   [[3]]
             [,1]       [,2]       [,3]       [,4]
   [1,] 0.21080765 0.09796696 0.14128463 0.08171739
   [2,] 0.09796696 0.06083267 0.07166802 0.04816533
   [3,] 0.14128463 0.07166802 0.11360874 0.07217753
   [4,] 0.08171739 0.04816533 0.07217753 0.06425322

   Component skewness matrices:
   [[1]]
              [,1]        [,2]        [,3]       [,4]
   [1,]  0.31881607 -0.27858936 0.007451723 0.12347998
   [2,]  0.07624952 -0.11956708 0.069916382 0.16058854
   [3,] -0.04518459 -0.15517472 0.171888937 0.07288276
   [4,]  0.01427354 -0.01726205 0.004160663 0.14294473

   [[2]]
              [,1]        [,2]        [,3]        [,4]
   [1,] -0.26806780  0.08357397 -0.5825049 0.254300197
   [2,]  0.15339662 -0.23236022 -0.3423885 0.045900090
   [3,]  0.05249894  0.22566719 -0.6678188 0.115936808
   [4,]  0.12559134  0.08309988 -0.2042694 0.007060828

   [[3]]
```

```
               [,1]          [,2]          [,3]          [,4]
[1,]   0.136541936  -0.13366639    0.51628869    0.3692744
[2,]  -0.017338343   0.15012982   -0.15683670    0.2290092
[3,]  -0.219878985   0.03899513    0.53218928    0.3051550
[4,]  -0.006770662   0.10861986   -0.04394092   -0.1217400

  Component degrees of freedom:
  51.74962 185.87390 126.98517

  Component mixing proportions:
  0.3333333 0.3314298 0.3352368
```

## 4.4. Nested special cases of the FM-CFUST distribution

In this section, we focus on the restricted and unrestricted versions of MST mixture models. For the normal and $t$-mixture models, routines for fitting them are implemented in **EM-MIXskew**. As noted earlier, the rMST distribution corresponds to a CFUST distribution with $q = 1$. Thus setting `q=1` in `fmcfust` will fit a FM-rMST model. However, as **EM-MIXskew** uses a specialized implementation of the EM algorithm for this model, the user is encouraged to use this package when fitting a FM-rMST model. Similarly, the **EMMIXuskew** package can be used for the fitting of a FM-uMST model. To fit the FM-rMST model to the same dataset as in the example above using the EMMIXcskew package, the following code can be used.

```
> fit.restricted <- fmcfust(3, iris[,-5],q=1)
```

The above model can also be fitted using the **EMMIXskew** package with the command `fit.restricted <- EmSkew(iris[,-5],3,"mst",debug=FALSE)`. We can compare the predicted cluster labels of these models against the true class labels. For all three models, the predicted cluster labels are stored as `clust` in the output object. A cross-tabulation of these labels suggests that the fitted FM-CFUST model performs well with only one misclassified observation, whereas the FM-rMST and FM-uMST models have three and six misclassified observations, respectively.

```
R> table(iris$Species, fit.iris$clust)

             1  2  3
  setosa     0 50  0
  versicolor 49  0  1
  virginica  0  0 50
R> table(iris$Species, fit.restricted$clust)

             1  2  3
  setosa     0  6 44
  versicolor 50  0  0
  virginica  50  0  0
```

| Model | FM-CFUST | FM-rMST | FM-uMST |
|-------|----------|---------|---------|
| **MCR** | 0.0067 | 0.3733 | 0.0200 |

Table 2: Misclassification rate of the three skew *t*-mixture models fitted to the `iris` dataset.

```
R> table(iris$Species, fit.unrestricted$clust)

            1  2  3
  setosa   50  0  0
  versicolor 0 47  3
  virginica  0  0 50
```

The misclassification rate (MCR) against the true labels can be calculated using `error.rate` from the **EMMIXskew** package. In this example, the FM-CFUST model obtained the lowest MCR compared to the FM-rMST and FM-uMST models (see Table 2 and the code below). Figure 2 shows the clustering of the `iris` dataset using these three models.

```
R> error.rate(unclass(iris$Species), fit.iris$clust)
[1] 0.006666667
R> error.rate(unclass(iris$Species), fit.restricetd$clust)
[1] 0.3733333
R> error.rate(unclass(iris$Species), fit.unrestricted$clust)
[1] 0.02
R>
R> panel1 <- function(x,y,...){points(x,y,col=c("red","green3","blue")
+ [fit.iris$clust],pch=20)}
R> panel2 <- function(x,y,...){points(x,y,col=c("red","green3","blue")
+ [fit.unrestricted$clust],pch=20)}
R> panel3 <- function(x,y,...){points(x,y,col=c("red","green3","blue")
+ [fit.restricted$clust],pch=20)}
R> pairs(iris[1:4], main = "Iris Data", pch = 20, col = c("red","green3","blue")
+ [unclass(iris$Species)], lower.panel=panel1)
R> pairs(iris[1:4], main = "Iris Data", upper.panel=panel2, lower.panel=panel3)
```

In the case of univariate data, note that all three models become identical, and thus the use of `EmSkew` from the **EMMIXskew** package is recommended as it provides a more computationally efficient implementation. It should be noted that the fitting of the FM-uMST and FM-CFUST models can be time consuming due to the amount of calculations required, especially when *q* is large. When tested on a 3.4GHz machine, the example in Section 4.2 took around 3.5 seconds to complete. For the examples in this Section, the CPU time for the FM-CFUST, FM-uMST, and FM-rMST models is around 1856, 1927, and 13 seconds, respectively. Note that if the specialised implementation of the **EMMIXskew** package is used, the CPU time for the FM-rMST model in this example reduces to 0.8 seconds.

# 5. Miscellaneous functions

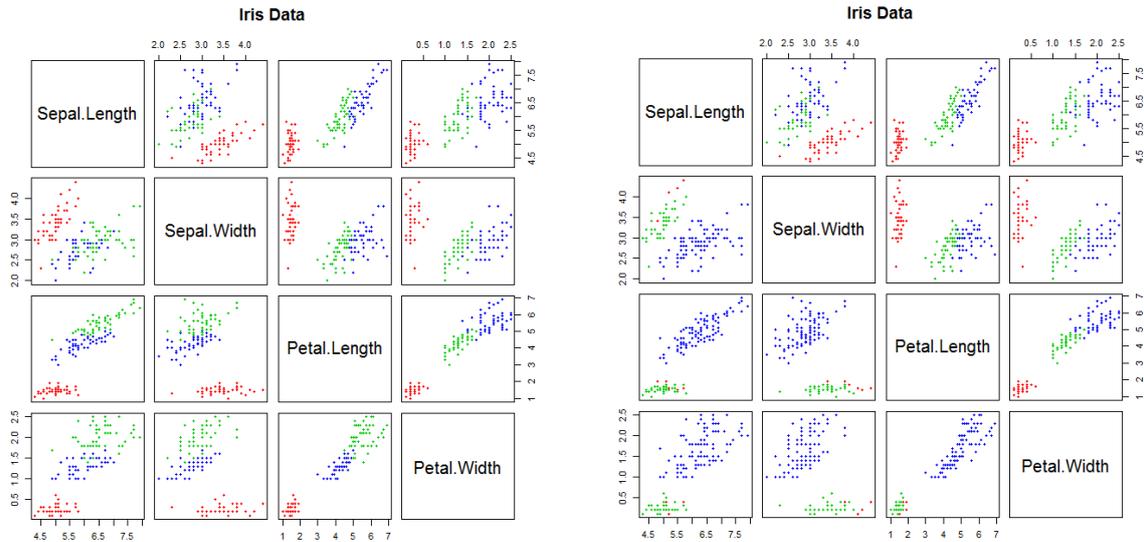

Figure 2: Clustering of the `iris` dataset. The upper panels of the figure on the left show the true labels, whereas the lower panels are the predicted labels given by the FM-CFUST model. For the figure on the right, the upper panels correspond to the results given by the FM-uMST model and the lower panels correspond to that given by the FM-rMST model.

This section presents some illustrative examples on how to generate random observations, generate/provide initial values (for the EM algorithm), use different stopping criteria, and draw contour plots for a FM-CFUST model using **EMMIXcskew**.

## 5.1. Random sample from the FM-CFUST distribution

The CFUST admits a convolution-type stochastic representation that facilitates random sample generation. More specifically, let $\boldsymbol{U}_0$ and $\boldsymbol{U}_1$ be independent random variables following multivariate normal distributions given by $N_p(\boldsymbol{0}, \boldsymbol{\Sigma})$ and $N_q(\boldsymbol{0}, \boldsymbol{I}_q)$, respectively. Let also $w$ be a scalar random variable with the gamma$(\frac{\nu}{2}, \frac{\nu}{2})$ distribution. Then

$$\boldsymbol{Y} = \boldsymbol{\mu} + \frac{1}{\sqrt{w}} \boldsymbol{\Delta} |\boldsymbol{U}_0| + \frac{1}{\sqrt{w}} \boldsymbol{U}_0 \tag{22}$$

has a CFUST distribution with density given by (1). The `rcfust` function adopts (22) to generate a random sample of CFUST observations. Its mixture version is implemented as `rfmcfust` in **EMMIXcskew**. These two functions are given, respectively, by

```
rcfust(n, mu, sigma, delta, dof, known=NULL, ...)
rfmcfust(g, n, mu, sigma, delta, dof, pro, known=NULL, ...)
```

Input arguments for the above functions follow the same structure as described in Section 4.1, permitting the parameters of the CFUST (or FM-CFUST) model to be specified either individually using `mu`, `sigma`, `delta`, and `dof` (and also `pro` for a FM-CFUST model) or within a `list` object using `known`. The argument `n` specifies the number of random observations to be generated. In the case of a FM-CFUST model, `n` is either a single integer (which represents the total number of observations to be generated) or a vector of `g` integers representing the number of observations to be generated from each of the `g` component. Note that if `n` is a single value, `rfmcfust` will determine the sample size for each component using the mixing proportion specified by `pro`.

As an example, suppose one would like to generate 10 random observations from a CFUST distribution with $\boldsymbol{\mu} = (1,2)^{\top}$, $\boldsymbol{\Sigma} = \boldsymbol{I}_2$, $\boldsymbol{\Delta} = \begin{bmatrix} 2 & 1 \\ 1 & 2 \end{bmatrix}$, and $\nu = 4$, the following command can be issued at the R command prompt. A $10 \times 2$ matrix will be returned.

```
R> rfust(10,c(1,2),diag(2),matrix(c(2,1,1,2),2,2),4)
           [,1]      [,2]
 [1,]  5.836001  5.600793
 [2,]  3.080172  4.213700
 [3,]  3.305617  4.888012
 [4,]  4.390739  3.109635
 [5,]  4.003996  4.686407
 [6,]  1.609795  1.599386
 [7,]  3.361534  5.326190
 [8,]  3.449745  4.474217
 [9,] 10.886028  7.964134
[10,]  5.752894  7.049037
```

Generating observations from a mixture of CFUST distributions is also quite simple using `rfmcfust`. We first create an object with the required parameters. This can then be directly passed in to `rfmcfust`. An example is shown below. A $n \times (p+1)$ matrix is returned where the last column gives component label for each data point generated.

```
R> obj <- list()
R> obj$mu <- list(c(17,19), c(5,22), c(6,10))
R> obj$sigma <- list(diag(2), matrix(c(2,0,0,1),2),
+ matrix(c(3,7,7,24),2))
R> obj$delta <- list(matrix(c(3,0,2,1.5),2,2), matrix(c(5,0,0,10),2,2),
+ matrix(c(2,0,5,0),2,2))
R> obj$dof <- c(1, 2, 3)
R> obj$pro <- c(0.25, 0.25, 0.5)
R> rfmcfust(3, 100, known=obj)

            [,1]         [,2] [,3]
 [1,] 46.143907 25.56151304    1
 [2,] 17.816665 18.22572581    1
 [3,] 33.915805 25.54308697    1
 [4,] 44.609637 15.81099978    1
 [5,] 54.766995 16.10253015    1
 [6,] 18.610320 19.74165026    1
 [7,] 25.303312 20.80981782    1
 [8,] 20.608770 21.58460735    1
 [9,] 19.679756 20.51390429    1
```

```
[10,] 20.988970 16.10998335     1
      ... the rest omitted ...
```

## 5.2. Starting values for fitting FM-CFUST distributions

Three different strategies for generating starting values for the FM-CFUST model were described in Section 3.3. These are implemented in the **EMMIXcskew** package. Apart from the default starting strategy (13) which makes use of moment-based estimates of a uMSN distribution, the `init.cfust` function implements the transformation approach (15) as one of its options, and accepts starting values based on the results of its nested models as another option. The arguments of the function are the following:

```
init.cfust(g, dat, q=p, initial=NULL, known=NULL, clust=NULL, nkmeans=20,
method=c("moments","transformation","EMMIXskew","EMMIXuskew"))
```

To use a fitted model given by the packages **EMMIXskew** and **EMMIXuskew**, set `method` to `"EMMIXskew"` and `"EMMIXuskew"`, respectively, and the output of the functions `EmSkew` and `fmmst` can be directly passed into `init.fust` using the argument `initial`. If an initial value of the parameter vector is not supplied (that is, `initial=NULL`), then the default option is to provide an initial value for each component-parameter vector obtained by applying the method of moments (that is, `method="moments"`) to the clusters corresponding to the components. These clusters are obtained by using the $k$-means procedure, but the user can specify an initial partition obtained using some other method of clustering, for example, a model-based approach using a mixture of $t$-distributions. The user-specified initial partition is passed in using `clust`. If the transformation approach is preferred, set `method="transformation"`. Again, in this case, if an initial partition is not supplied, the partition given by $k$-means clustering is used.

An example session is shown below demonstrating how to use the above function to generate different starting values. We use the Geyser dataset (Azzalini and Bowman 1990), which contain measurements on 299 successive eruptions of the Old Faithful Geyser during 1 August to 15 August 1985. The two variables recorded were the waiting time between eruptions and the duration of each eruption, both measured in minutes. This dataset is available from the **MASS** package.

```
R> library(MASS)
R> data(geyser)
R> plot(geyser, pch=20)
```

An initial inspection of the dataset (Figure 3(a)) suggests three clusters. Hence, we set $g$ to 3. In the example below, `initial.default` and `initial.transformation` refers to default (moment-based) approach and the transformation approach, respectively. For the nested approach, we have demonstrated in Section 4.3 how to use the results of a fitted FM-uMST model as initial values. In that example, the model was fitted using `fmmst()` in our package, which is a replica of the same function in the **EMMIXuskew** package, and hence the option `method="EMMIXuskew"` was used. This option can be used in the same way to supply initial values from the **EMMIXuskew** package. In addition, the **EMMIXcskew** package also accepts outputs from the **EMMIXskew** package which provide routines to fit finite mixtures of normal, *t*, (restricted) skew-normal, and (restricted) skew *t*-distributions. In this case, the option `method="EMMIXskew"` is used to pass the results to `fmcfust()`.

```
R> initial.default <- init.cfust(3, geyser)
R> initial.transformation <- init.cfust(3, geyser, method="transformation")
R> fit.geyser.restricted <- EmSkew(geyser, 3, "mst", debug=FALSE)
R> initial.restricted <- init.cfust(3, geyser, initial=fit.gesyser.restricted,
+ method="EMMIXskew")
R> fit.geyser.unrestricted <- fmmst(3, geyser)
R> initial.unrestricted <- init.cfust(3, geyser, initial=fit.gesyser.unrestricted,
+ method="EMMIXuskew")
R> fit.geyser.t <- EmSkew(geyser, 3, "mvt", debug=FALSE)
R> initial.t <- init.cfust(3, geyser, initial=fit.gesyser.t, method="EMMIXskew")
```

To help choose an appropriate starting value, we can compare the log likelihood values for the FM-CFUST model fitted using these initial values.

```
R> initial.default$loglik
[1] -1903.598
R> initial.transformation$loglik
[1] -1448.33
R> initial.restricted$loglik
[1] -1335.038
R> initial.unrestricted$loglik
[1] -1404.772
R> initial.t$loglik
[1] -1347.385
```

In this case, the starting value that corresponds to the fitted model of FM-rMST gave the highest log likelihood value. We now proceed to fit a FM-CFUST model using the default starting strategy, the transformation approach, and the fitted model of FM-rMST.

```
R> fit.geyser1 <- fmcfust(3, geyser, initial=initial.default)
R> fit.geyser2 <- fmcfust(3, geyser, initial=initial.transformation)
R> fit.geyser3 <- fmcfust(3, geyser, initial=initial.restricted)
R> fit.geyser1$loglik
[1] -1415.442
R> fit.geyser2$loglik
[1] -1344.724
R> fit.geyser3$loglik
[1] -1333.533
```

According to the use of the final log likelihood value, the results of `fit.geyser3` are preferred. This corresponds to the result using the initial value with the highest log likelihood value identified above. Note that this may not always be the case; that is, using the initial value with the highest log likelihood value may not always give the optimal results compared to those with smaller initial log likelihood values. It is advisable to run the EM algorithm using a range of different starting values. To visualize the clustering results of the above three models, we can plot the data with colours according to the predicted cluster labels given by these models, as shown below. Figures 3(a), 3(b), and 3(c) show the results using the default strategy, the transformation approach, and the fitted FM-rMST model, respectively. It can be observed that `fit.geyser2` perhaps gave a more natural partition of the data, although its log likelihood value is lower than that given by `fit.geyser3`.

```
R> plot(geyser, pch=20, col=c("red","blue","green")[fit.geyser1$clust])
R> plot(geyser, pch=20, col=c("red","blue","green")[fit.geyser2$clust])
R> plot(geyser, pch=20, col=c("red","blue","green")[fit.geyser3$clust])
```

## 5.3. Stopping Criteria

The stopping criteria described in Section 3.4 are available through the `convergence` option in `fmcfust()`. The default is using Aitken's acceleration-based approach (`convergence="Aitken"`). The other two options are `convergence="likelihood"` and `convergence="parameters"`, referring to relative likelihood based and relative parameters-based approach respectively. For illustration, using the Geyser dataset and `intial.restricted` as initial values as an example, we can run the EM algorithm with the relative likelihood-based and relative parameter-based convergence criteria using the following commands.

```
R> fit.geyser4 <- fmcfust(3, geyser, initial = initial.restricted,
+ convergence=c("likelihood"))

Finite Mixture of Multivariate CFUST Distributions
with  3 components
  ---------------------------------------------------

  Iteration  0 : loglik =  -1335.038
  Iteration  1 : loglik =  -1335.01
  Iteration  2 : loglik =  -1334.987
  Iteration  3 : loglik =  -1334.965
  Iteration  4 : loglik =  -1334.943
  Iteration  5 : loglik =  -1334.922
              ... rest omitted ...
  ---------------------------------------------------
  Iteration 100: loglik = -1333.533

R. fit.geyser5 <- fmcfust(3, geyser, initial = initial.restricted,
+ convergence=c("parameters"))

Finite Mixture of Multivariate CFUST Distributions
with  3 components
  ---------------------------------------------------
```

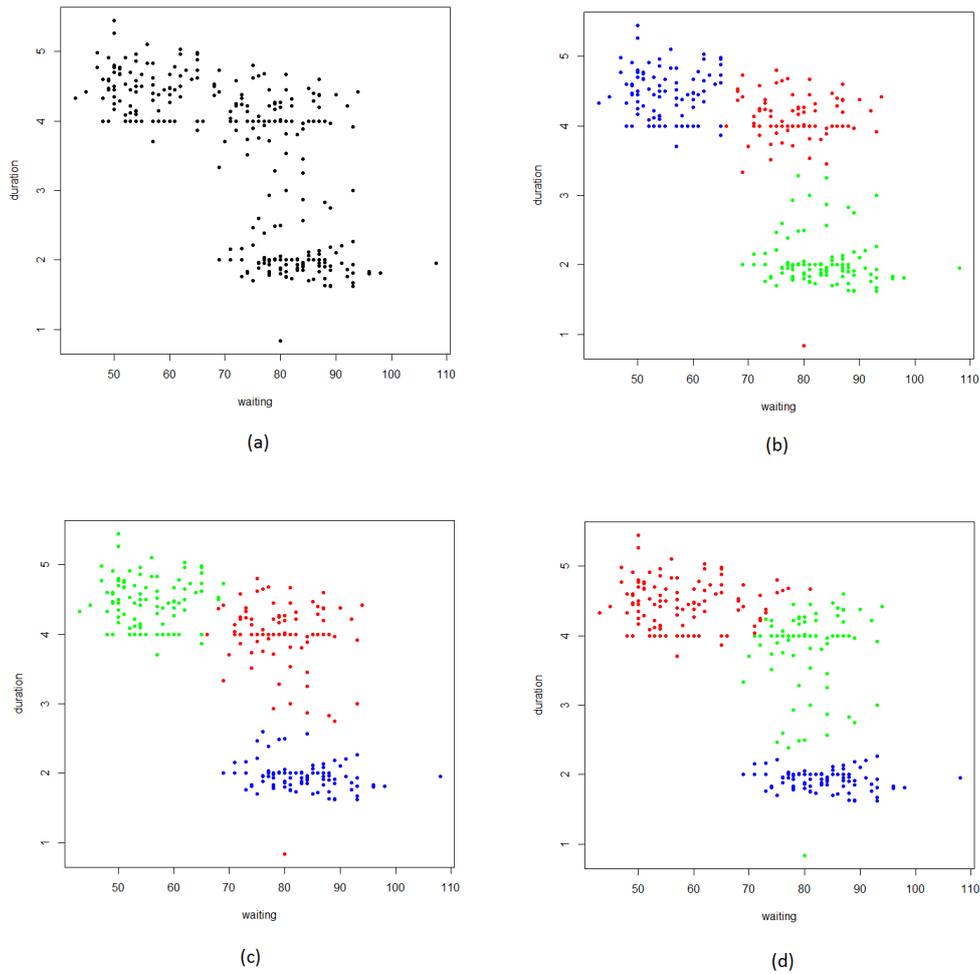

Figure 3: The Old Faithful geyser data from the **MASS** package. (a) Scatter plot of the data. (b) Clustering results of the FM-CFUST model using the default (moments-based) approach for generating starting values. (c) Clustering results of the FM-CFUST model using the transformation approach for generating starting values. (d) Clustering results of the FM-CFUST model using the results of FM-rMST model as starting values.

```
Iteration  0 : loglik =  -1335.038
Iteration  1 : loglik =  -1335.01
Iteration  2 : loglik =  -1334.987
Iteration  3 : loglik =  -1334.965
Iteration  4 : loglik =  -1334.943
Iteration  5 : loglik =  -1334.922
              ... rest omitted ...
---------------------------------------------------
Iteration 100: loglik = -1333.533
```

In both cases, we can observe from the output above that the EM algorithm terminates in the same iteration for this example.

### 5.4. Choosing the number of components $g$ with BIC

Model selection criteria are typically used to guide the choice of $g$ when fitting finite mixture models. The `fmcfust()` function provides the values of AIC and BIC as part of the output when fitting a FM-CFUST model. We show here a short example of using BIC to assist in choosing the optimal value of $g$. We fit the FM-CFUST model with $g = 1, \ldots, 4$ to the Geyser data, using the default starting strategy. In this case, the lowest BIC corresponds to the model with $g = 2$.

```
R> fit.geyser.g1 <- fmcfust(1, geyser)
R> fit.geyser.g2 <- fmcfust(2, geyser)
R> fit.geyser.g3 <- fit.geyser1
R> fit.geyser.g4 <- fmcfust(4, geyser)

R> fit.geyser.g1$bic
[1] 3166.698
R> fit.geyser.g2$bic
[1] 2963.833
R> fit.geyser.g3$bic
[1] 3013.298
R> fit.geyser.g4$bic
[1] 3069.547
```

### 5.5. Visualization of fitted contours

Contour plots for a FM-CFUST model can be produced easily using the functions `fmcfust.contour.2d` and `fmcfust.contour.3d` for a two-dimensional and three-dimensional space, respectively. These two functions take a number of arguments described below.

```
fmcfust.contour.2d(dat, model, grid=50, drawpoints=TRUE, clust=NULL,
    nlevels=10, component=NULL, ...)
fmcfust.contour.3d(dat, model, grid=20, drawpoints=TRUE, clust=NULL,
    levels=0.9, component=NULL, ...)
```

Briefly, `dat` is a dataset that is either a `matrix` or `data.frame`. Note that if `dat` is not specified, then the limits of the axes of the plot must be specified (using the standard `xlim`, `ylim`, and `zlim` arguments). The parameters of the FM-CFUST model are specified using `model`. Typically, this is an output from `fmcfust`. The argument `grid` determines the grid size of the plots. Thus the higher the number in `grid`, the smoother the contours (at the cost of longer computation time). The data points (if provided) are plotted by default. By setting `drawpoints=FALSE`, only the contours will be plotted. In the case where $g > 1$, a user-specified partition of the data can be provided using `clust`. This is used when `drawpoints=TRUE` and the data points in the plot will be colour-coded using the labels in `clust`. The arguments `nlevels` and `levels` control how many contours are displayed and at which percentiles are they computed, respectively. Finally, `component` specifies which component is included in the plot. This option allows the components to be plotted individually without taking into account the mixing proportions. In contrast, the default is to return the contours of a mixture density.

To illustrate the use of `fmcfust.contour.2d`, we reconsider the `iris.versicolor` example in Section 4.2. Figure 1 can be generated by the following command in R.

```
R> fmcfust.contour.2d(iris.versicolor, fit.Versicolor, drawpoints=FALSE,
+   main="versicolor")
```

We now turn to an example of generating a 3D contour plot of a FM-CFUST model. Suppose we would like to draw the contours of a two-component FM-CFUST distribution with parameters $\boldsymbol{\mu}_1 = (0, 0, 0)^\top$, $\boldsymbol{\mu}_2 = (5, 5, 5)^\top$, $\nu_1 = \nu_2 = 3$, $\pi = (0.2, 0.8)$,

$$\boldsymbol{\Sigma}_1 = \begin{bmatrix} 5 & 2 & 1 \\ 2 & 5 & 1 \\ 1 & 1 & 1 \end{bmatrix}, \ \boldsymbol{\Sigma}_2 = 2I_3, \ \boldsymbol{\Delta}_1 = \begin{bmatrix} 1 & 0 & 0 \\ 1 & 0 & 0 \\ 1 & 0 & 0 \end{bmatrix}, \text{ and } \boldsymbol{\Delta}_2 = \begin{bmatrix} 5 & 0 & 0 \\ 0 & 10 & 0 \\ 0 & 0 & 15 \end{bmatrix}.$$

We first create an object `obj` with these parameters. By default, a mixture density is produced on running the `fmcfust.contour.3d` (see Figure 4(a)). A first remark on this figure is that the two components seem to be 'joined' together. To gain a better view of the shapes of these two components, we may set `components=1:2` so that their mixing proportions are ignored. In addition, we generate 500 random observations from the specified FM-CFUST model and include them in the plot. Observe now in Figure 4(b) that the two components are plotted as two separately objects and their colours are matched with the simulated data.

```
R> obj <- list()
R> obj$mu <- list(matrix(c(0,0,0),3), matrix(c(5,5,5),3))
R> obj$sigma <- list(matrix(c(5,2,1,2,5,1,1,1,1),3,3), 2*diag(3))
R> obj$delta <- list(matrix(c(1,0,0,1,0,0,1,0,0),3,3),
+   matrix(c(5,0,0,0,10,0,0,0,15),3,3))
R> obj$dof <- c(3,3)
R> obj$pro <- c(0.2, 0.8)
R> fmcfust.contour.3d(model=obj, level=0.98, drawpoints=TRUE, xlab="X",
+   ylab="Y", zlab="Z", xlim=c(-20, 50), ylim=c(-20,50), zlim=c(-20,80))
R> X <- rfmcfust(2, 500, known=obj)
R> fmcfust.contour.3d(X, model=obj, level=c(0.99, 0.92), drawpoints=TRUE,
+   clust=X[,4], xlab="X", ylab="Y", zlab="Z", xlim=c(-20, 50), ylim=c(-20, 50),
+   zlim=c(-20, 80), component=1:2)
```

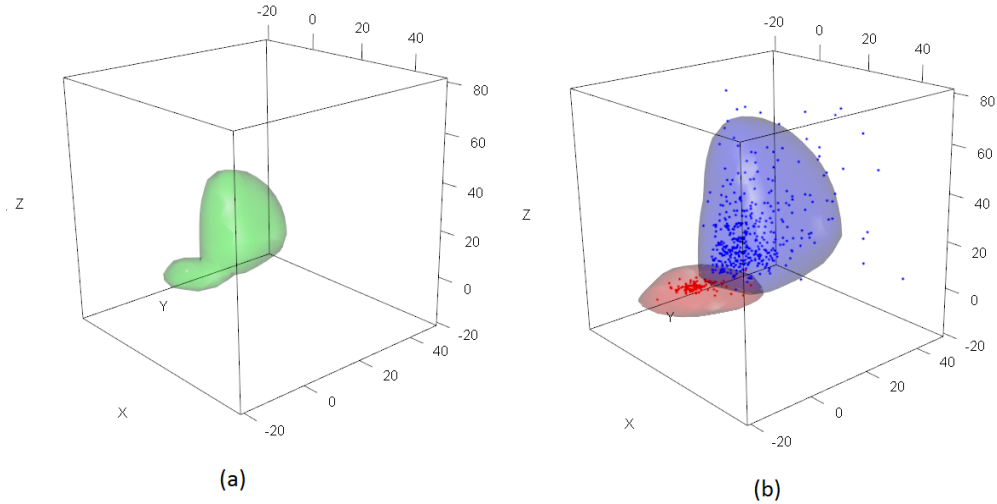

Figure 4: Contour plots of FM-CFUST models generated by `EMMIXcskew` package. (a) 3D contours of the density of a FM-CFUST distribution. (b) 3D contours of the density of each component of the FM-CFUST distribution shown in (b).

## 6. Concluding remarks

This paper has presented the **EMMIXcskew** package for the fitting of a CFUST distribution and finite mixtures of CFUST distributions to heterogeneous data that exhibit non-normal features. In addition to computing the maximum likelihood estimates of the model parameters, the **EMMIXcskew** package provides routines for random number generation, density evaluation, the plotting of 2D and 3D contours, and a few different methods for initial values generation for the FM-CFUST model. A finite mixture of CFUST distributions provides a model for the robust extension of traditional normal mixture models, with greater flexibility in handling asymmetry and heavy tails. The skewness parameters of a CFUST distribution are characterized by a general matrix, which provides an elegant unification of the restricted and unrestricted skew $t$-distributions. The aim of this package is to provide users with the option of fitting this flexible distribution to their dataset. Model selection criteria such as the AIC and BIC are provided for the FM-CFUST model to assist the user in choosing between different models for their data.

It is noted that the fitting of a FM-CFUST model can be quite computationally intensive when $q$ is large. This is due to the calculations of some of the conditional expectations involved in the E-step of the EM algorithm. Future work will consider applicable strategies to speed up the parameter estimation process for this model such as parallel implementations described in Lee *et al.* (2016b,a) and Lee and McLachlan (2016b).

## Acknowledgments

This work is supported by grants from the Australian Research Council.

**Affiliation:**

Sharon X. Lee
School of Mathematics and Physics
University of Queensland
Brisbane, Australia
E-mail: s.lee11@uq.edu.au

Geoffrey J. McLachlan
School of Mathematics and Physics
University of Queensland
Brisbane, Australia
E-mail: g.mclachlan@uq.edu.au